\documentstyle[12pt,a4wide]{article}
\newcommand{\nl}{ {\hfill \break} }
\newcommand{\np}{ {\newpage } }

\newcommand{\R}{ \mbox{\rm I$\!$R} }

\newcommand{\sign}{ \mbox{\rm sign} }

\newcommand{\artanh}{ \mbox{\rm artanh} }
\newcommand{\arcoth}{ \mbox{\rm arcoth} }

\newcommand{\be}[1]{\begin{equation}\label{#1}}
\newcommand{\ee}{\end{equation}}
\newcommand{\ba}[1]{\begin{eqnarray}\label{#1}}
\newcommand{\ea}{\end{eqnarray}}

\begin{document}

\bigskip

\bigskip

\vspace{-0.2truecm}

\centerline{\large \bf A POSSIBLE SOLUTION TO}
\centerline{\large \bf THE PROBLEM OF EXTRA DIMENSIONS}

\vspace{1.03truecm}

\bigskip

\centerline{\bf \large U. Bleyer\dag, M. Rainer\dag, A. Zhuk{\dag \ddag}}

\vspace{0.96truecm}

\centerline{\dag Gravitationsprojekt, Universit\"at Potsdam}
\centerline{An der Sternwarte 16}
\centerline{D-14482 Potsdam, Germany}

\vspace{0.15truecm}

\centerline{\ddag Fachbereich Physik, Freie Universit\"at Berlin}
\centerline{Arnimallee 14}
\centerline{ D-14195 Berlin, Germany \footnote{Permanent address:
Department of Physics,
University of Odessa,
2 Petra Velikogo,
Odessa 270100, Ukraine}}

\vspace{3.35truecm}

\centerline{\bf Abstract}

\vspace{1.07truecm}

{\small
\noindent
We consider a multidimensional universe with the topology
$M= \R\times M_1\times \cdots \times M_n$,
where the $M_i$ ($i>1$) are $d_i$-dimensional Ricci flat spaces.
Exploiting a conformal equivalence between minimal
coupling models and conformal coupling models,
we get exact solutions for such an universe filled by
a conformally coupled scalar field.
One of the solutions can be used to describe trapped unobservable extra
dimensions.



\np
\section{\bf Introduction}
\setcounter{equation}{0}
Recently models of multidimensional universes
$M= \R\times M_1\times \cdots \times M_n$, where
$M_i$, ($i=1,\ldots,n$) are Einstein spaces,
have received increasing interest \cite{IMZ}.
The geometry might be minimally coupled to
a homogeneous scalar field $\Phi$ with a potential
$U(\Phi)$.
The class of multidimensional cosmological models (MCM) is rich
enough to study the relation and the imprint of internal compactified
extra dimensions (like in  Kaluza-Klein models \cite{aBlLP,bBlLP}) on
the external space-time.
Therein, exactly solvable classical and quantum models were found
by \cite{IZ}.
Some of these exact solutions describe the compactification of the internal
dimensions up to the actual time.
Accordingly, all MCM can be divided into two different classes:
The first class consists of models where from the very beginning the internal
dimensions are assumed to be static with a scale of Planck length
$L_{Pl}\sim 10^{-33}\mbox{cm}$ \cite{aBZ,bBZ}.
The other class consists of models where the internal dimensions,
like external space-time, evolve dynamically.
Thereby however, the internal spaces contract for several orders
of magnitude relatively to the external one \cite{aBZ,cBZ}.

In \cite{IZ}-\cite{cBZ} a minimally coupled scalar field as a matter source
was considered. More in particular, in \cite{dBZ} a generalized
Kasner solution was found, with minimally coupled scalar field
and all spaces $M_i$ being Ricci flat.
Maeda \cite{Ma} (see also \cite{Xa}) has shown the equivalence
of a minimally coupled model to a model with arbitrary coupled
scalar field. In $4$ dimensions, this was used by Page \cite{Pa}
to get new solutions with conformally coupled scalar field from some
known solutions with minimally coupled scalar field.
This idea can be exploited also, more generally,
for arbitrary dimensions and coupling constants \cite{aRa}.
In the present paper, from the multidimensional solution of \cite{dBZ},
we obtain a new solution with conformally coupled scalar field.
This new solution bears the possibility to solve the problem
of extra dimensions.
Rubakov and Shaposhnikov \cite{RS} and others \cite{Vis}-\cite{Mai}
consider non-gravitational fields as dynamical variables which are
trapped by a potential well (domain wall),
which is narrow along corresponding
internal dimensions and extended flatly in $4$-dimensional space-time.
In \cite{Am} also dynamical variables of gravity are trapped by a
potential. We propose that some gravitational degrees of freedom may be
trapped in a classically forbidden region and, hence, be invisible.
In general the scale factors of all factor spaces $M_i$ are
dynamical variables of the MCM.
Here we obtain solutions where the internal spaces are still
trapped invisibly, while external $3$-space is already born.
The factor spaces (external or internal) are born in a quantum
tunnelling process from ``nothing'' \cite{Tr}-\cite{Vi}, i.e.
from a non-real (e.g. imaginary) section in complex geometry.
The birth of different factor spaces may happen at different times.
Some of them may remain confined forever in a non-classical section.
In complex geometry the extra dimensions also correspond to
resolutions of simple singularities in a $3+1$-dimensional
space-time,
in which their real parts evolve as string tubes \cite{bRa}.
In the following, we explore these ideas in the example of a new solution.

\section{\bf Classical Multidimensional Universes}
\setcounter{equation}{0}
We consider a universe described by a (Pseudo-) Riemannian manifold
$$
M=\R\times M_1 \times\ldots\times M_n,
$$
with first fundamental form
\begin{equation}
g\equiv ds^2 = -e^{2\gamma} dt\otimes dt
     + \sum_{i=1}^{n} a_i^2 \, ds_i^2,
\end{equation}
where    $ a_i=e^{\beta^i} $
is the scale factor of the $d_i$-dimensional space $M_i$.
In the following we assume $M_i$ to be an Einstein space,
i.e. its first fundamental form
\begin{equation}
ds_i^2
=g^{(i)}_{{k}{l}}\,dx^{k}_{(i)} \otimes dx^{l}_{(i)}
\end{equation}
satisfies the equations
\begin{equation}
R^{(i)}_{kl}=\lambda_i g^{(i)}_{kl},
\end{equation}
and hence
\begin{equation}
R^{(i)}=\lambda_i d_i.
\end{equation}
For the metric (2.1) the Ricci scalar curvature of $M$ is
\begin{equation}
R=e^{-2\gamma}\left \{
\left [ \sum_{i=1}^{n} (d_i \dot\beta^i) \right ]^2
+ \sum_{i=1}^{n}
d_i[ {(\dot\beta^i)^2-2\dot\gamma\dot\beta^i+2\ddot \beta^i} ]
\right \}
+\sum_{i=1}^{n} R^{(i)} e^{-2\beta^i}.
\end{equation}

Let us now consider a variation principle with the action
\be{2.10}
S=S_{EH}+S_{GH}+S_{M},
\ee
where
\[
S_{EH}=\frac{1}{2\kappa^2}\int_{M}\sqrt{\vert g\vert} R\, dx
\]
is the Einstein-Hilbert action,
\[
S_{GH}=\frac{1}{\kappa^2}\int_{\partial M}\sqrt{\vert h\vert} K \, dy
\]
is the Gibbons-Hawking boundary term \cite{Gi}, where $K$  is the trace
of the second fundamental form, which just cancels
second time derivatives in the equation of motion,
and
\[
S_{M}=\int_{M}\sqrt{\vert g\vert}
[-\frac{1}{2}g^{ik}\partial_i\Phi\partial_j\Phi-U(\Phi)]\, dx
\]
is a matter term. Here, and in the following, $\Phi$ is a
homogeneous minimally coupled scalar field.
In the case of minimal coupling, we denote the
lapse function by $e^{\hat\gamma^i}$, and the other
scale factors by $\hat a_i\equiv e^{\hat\beta^i}$.
Then we define the metric on minisuperspace, given in
the coordinates $\hat\beta^i$ and $\Phi$. We set
\begin{equation}
G_{ij}:=d_i \delta_{ij}-d_i d_j
\end{equation}
and define the minisuperspace metric as
\begin{equation}
G = G_{ij}d\hat\beta^i\otimes d\hat\beta^j
+\kappa^2 d\Phi \otimes d\Phi.
\end{equation}
Furthermore we define
\begin{equation}
N:=e^{\hat\gamma-\sum_{i=1}^{n}d_i\hat\beta^i}
\end{equation}
and a minisuperspace potential $V=V(\hat\beta^i,\Phi)$ via
\begin{equation}
V:=-\frac{\mu}{2}\sum_{i=1}^{n}R^{(i)}
e^{-2\hat\beta^i+\hat\gamma+\sum_{j=1}^{n}d_j\hat\beta^j}
+\mu\kappa^2U(\Phi)e^{\hat\gamma+\sum_{j=1}^{n}d_j\hat\beta^j},
\end{equation}
where
\begin{equation}
\mu:=\kappa^{-2}\prod_{i=1}^{n}\sqrt{\vert\det g^{(i)}\vert}.
\end{equation}
Then the
variational principle of (\ref{2.10}) is  equivalent to
a Lagrangian variational principle in minisuperspace,
\be{2.16}
S=\int Ldt, \quad{\rm where}\quad
L=N\{\frac{\mu}{2}N^{-2}
(G_{ij}\dot{\hat\beta^i}\dot{\hat\beta^j}+\kappa^2\dot\Phi^2)
-V\}.
\ee
Here $\mu$ is the mass of a classical particle in minisuperspace.
Note that $\mu^2$ is proportional to the volumes of spaces $M_i$.

Next let us compare different choices of time $\tau$  in Eq. (2.1).
The time gauge is determined by the function $\gamma$.
There exist few natural time gauges (compare also \cite{aRa}).
In the following we need only:

i) The {\em synchronous time gauge}
\begin{equation}
\gamma\equiv 0,
\end{equation}
for which $t$ in Eq. (2.1) is the proper time $t_s$ of the universe.
The clocks of geodesically comoved observers go synchronous to that
time.

ii) The {\em harmonic time gauge}
\be{2.27}
\gamma\equiv\gamma_h:=\sum_{i=1}^n d_i\beta^i
\ee
yields the time $t\equiv t_h$, given by
\begin{equation}
dt_h=\left( \prod_{i=1}^n a_i^{d_i} \right)^{-1} dt_s
\end{equation}
In this gauge the time
is a harmonic function, i.e. $\Delta[g] t =0$, and $N\equiv 1$.
The latter is especially convenient when we work in
minisuperspace.

In the harmonic time gauge
the equations of motion from Eq. (\ref{2.16}) yield
\be{e2.16}
\mu G_{ij}\ddot{\hat\beta^j}=-\frac{\partial V}{\partial{\hat\beta^i}}
\qquad
\ddot\Phi+\frac{\partial U}{\partial\Phi}e^{2\hat\gamma}=0
\ee
plus the energy constraint
\be{e2.17}
\frac{\mu}{2}
(G_{ij}\dot{\hat\beta^i}\dot{\hat\beta^j}+\kappa^2\dot\Phi^2)+V=0.
\ee

\section{\bf Conformally Related Models}
\setcounter{equation}{0}

Let us follow \cite{Ma}  and consider an action of the kind
\begin{equation}
S=\int d^Dx\sqrt{\vert g\vert}(F(\phi,R)-\frac{\epsilon}{2}(\nabla\phi)^2).
\end{equation}
With
\begin{equation}
\omega:=\frac{1}{D-2}\ln(2\kappa^2
			  \vert\frac{\partial F}{\partial R}\vert)+A,
\end{equation}
where $D=1+\sum_{i=1}^{n}d_i$ and $A$ is an arbitrary constant,
$g_{\mu\nu}$ is conformally transformed to the minimal metric
\begin{equation}
\hat g_{\mu\nu}=e^{2\omega}g_{\mu\nu}.
\end{equation}

Especially let us consider in the following actions, which are linear
in $R$. With
\begin{equation}
F(\phi,R)=f(\phi)R-V(\phi).
\end{equation}
the action is
\begin{equation}
S=\int d^Dx\sqrt{\vert g\vert}(f(\phi)R-V(\phi)
-\frac{\epsilon}{2}(\nabla\phi)^2).
\end{equation}

In this case
\be{3.15}
\omega=\frac{1}{D-2}\ln(2\kappa^2
\vert f(\phi)\vert)+A
\ee
The scalar field in the minimal model is
$$
\Phi=\kappa^{-1}\int d\phi\{\frac{\epsilon(D-2)f(\phi)+2(D-1)(f'(\phi))^2}
			       {2(D-2)f^2(\phi)}\}^{1/2}  =
$$
\begin{equation}
=(2\kappa)^{-1}\int d\phi\{\frac{2\epsilon f(\phi)+\xi_c^{-1}(f'(\phi))^2}
			       {f^2(\phi)}\}^{1/2},
\end{equation}
where
\begin{equation}
\xi_c:=\frac{D-2}{4(D-1)}
\end{equation}
is the conformal coupling constant.

For the following we define $\sign x$ to be $\pm 1$ for $x\geq 0$ resp. $x<0$.
Then with the new minimally coupled potential
\be{3.18}
U(\Phi)=(\sign f(\phi))\ [2\kappa^2\vert f(\phi)\vert]^{-D/D-2}V(\phi)
\ee
the corresponding minimal action is
\begin{equation}
S=\sign f\int d^Dx\sqrt{\vert \hat{g}\vert}\left(-\frac{1}{2}
	  [(\hat{\nabla}\Phi)^2-\frac{1}{\kappa^2}\hat{R}]-U(\Phi)\right).
\end{equation}

Example 1:
\begin{equation}
f(\phi)=\frac{1}{2}\xi\phi^2,
\end{equation}
\begin{equation}
V(\phi)=-\lambda\phi^\frac{2D}{D-2}.
\end{equation}
Substituting this into Eq. (\ref{3.18}) the corresponding minimal potential
$U$ is constant,
\begin{equation}
U(\Phi)=(\sign \xi)\ \vert\xi\kappa^2\vert^{-D/D-2} \,\lambda.
\end{equation}
It becomes zero precisely for $\lambda=0$, i.e. when $V$ is zero.
With
\begin{equation}
f'(\phi)=\xi\phi
\end{equation}
we obtain
$$
\Phi=\kappa^{-1}\int d\phi
\left\{
\frac{ ({\epsilon\over\xi} + {1\over{\xi_c}} ) \phi^2}{\phi^4}
\right\}^\frac{1}{2}
=\left(\kappa\sqrt\xi\right)^{-1}
\sqrt{\frac{1}{\xi_c}+ {\epsilon\over\xi} }
\int d\phi\frac{1}{\vert\phi\vert}
$$
\begin{equation}
=\kappa^{-1}\sqrt{\frac{1}{\xi_c}+{\epsilon\over\xi} }\,\ln\vert\phi\vert
+k
\end{equation}
for $-\frac{\xi}{\epsilon}\geq\xi_c$,
where $k$ is a constant of integration.
Note that for
\begin{equation}
\frac{\xi}{\epsilon}=-\xi_c,
\end{equation}
e.g. for $\epsilon=-1$ and conformal coupling, we have
\begin{equation}
\Phi=k.
\end{equation}
Thus here the conformal coupling theory is equivalent to
a theory without scalar field.
For $-\frac{\xi}{\epsilon}<\xi_c$ the field $\Phi$ would become
complex and, for imaginary $k$, purely imaginary.
In any case, the integration constant $k$ may be a function of the
coupling $\xi$ and the dimension $D$.

Example 2:\nl
\be{3.27}
f(\phi)=\frac{1}{2}(1-\xi\phi^2),
\ee
\begin{equation}
V(\phi)=\Lambda.
\end{equation}
Then the constant potential $V$ has its minimal correspondence in a
non constant $U$, given by
\begin{equation}
U(\Phi)=\pm \Lambda \vert\kappa^2 (1-\xi\phi^2) \vert^{-D/D-2}
\end{equation}
respectively
for $\phi^2<\xi^{-1}$ or $\phi^2>\xi^{-1}$.

Let us set in the following
\begin{equation}
\epsilon=1.
\end{equation}
Then with
\begin{equation}
f'(\phi)=-\xi\phi
\end{equation}
we obtain
\be{3.32}
\Phi=\kappa^{-1}\int d\phi\{\frac{1+c\,\xi\phi^2}
			       {(1-\xi\phi^2)^2}\}^{1/2},
\ee
where
\begin{equation}
c:=\frac{\xi}{\xi_c}-1.
\end{equation}
For $\xi=0$ it is $\Phi=\kappa^{-1}\phi +k$, i.e. the coupling remains
minimal.
To solved this integral for $\xi\neq 0$, we substitute $u:=\xi\phi^2$.
To assure a solution of (\ref{3.32}) to be real, let us assume $\xi\geq\xi_c$
which yields $c\geq 0$.
Then we obtain
$$
\Phi=\frac{\sign(\phi)}{2\kappa\sqrt{\xi}}
		  \int{\frac {\sqrt {u^{-1}+c}}{\vert1-u\vert}}du
+k_{<\atop>}
$$
$$
=\frac{\sign((1-u)\phi)}{2\kappa\sqrt{\xi}}
[-\sqrt {c}\ln (2\,\sqrt {c}\sqrt {1+cu}\sqrt {u}+2\,cu+1)+
$$
$$
\sqrt {1+c}
\ln ({\frac{2\,\sqrt{1+c}\sqrt{1+cu}\sqrt {u}+2\,cu+1+u}{\vert1-u\vert}})]
+k_{<\atop>}
$$
$$
=\frac{\sign((1-\xi\phi^2)\phi)}{2\kappa\sqrt{\xi}}
\{-\sqrt {c}\ln (2\,\sqrt {c}\sqrt {1+c\xi\,\phi^{2}}
\sqrt {\xi} \vert\phi\vert
+2\,c\xi\,\phi^{2}+1)
$$
$$
+ \sqrt {1+c}\ln ({\frac {2\,\sqrt {1+c}
\sqrt{1+c\xi\,\phi^{2}} \sqrt {\xi} \vert\phi\vert
+2\,c\xi\,\phi^{2}+1+\xi\,\phi^{2}}
{\vert 1-\xi\,\phi^{2}\vert}})\}
+k_{<\atop>}
$$
$$
=\frac{\sign((1-\xi\phi^2)\phi)}{2\kappa\sqrt{\xi}}
\ln
\frac{ [2\,\sqrt {1+c} \sqrt{1+c\xi\,\phi^{2}}\sqrt {\xi}\vert\phi\vert
	    +(2\,c+1)\xi\,\phi^{2}+1]^{\sqrt {1+c}}  }
{ [2\,\sqrt {c}\sqrt {1+c\xi\,\phi^{2}}\sqrt {\xi}\vert\phi\vert
	     +2\,c\xi\,\phi^{2}+1]^{\sqrt{c}}
\cdot {\vert 1-\xi\,\phi^{2}\vert}^{\sqrt {1+c}}  }
$$
\be{3.34}
+k_{<\atop>}.
\ee
The integration constants $k_{<\atop>}$
for $\phi^2<\xi^{-1}$ and $\phi^2>\xi^{-1}$ respectively
may be arbitrary functions of $\xi$ and the dimension $D$.
The singularities of the transformation $\phi\to\Phi$ are located at
$\phi^2=\xi^{-1}$.

If the coupling is conformal $\xi=\xi_c$,
i.e. $c=0$, the expressions (\ref{3.34}) simplify to
\begin{equation}
\kappa\Phi=\frac{1}{\sqrt{\xi_c}} [(\artanh\sqrt{\xi_c}\phi)+c_<]
\end{equation}
for $\phi^2<\xi^{-1}_c$ and to
\begin{equation}
\kappa\Phi=\frac{1}{\sqrt{\xi_c}} [(\arcoth\sqrt{\xi_c}\phi)+c_>]
\end{equation}
for $\phi^2>\xi^{-1}_c$,
with redefined constants of integration $c_{<\atop>}$.
In the following we restrict to this case of conformal coupling.
The inverse formulas expressing the conformal field $\phi$ in terms of
the minimal field $\Phi$ are
\be{3.37}
\phi=\frac{1}{\sqrt{\xi_c}} \left[ \tanh(\sqrt{\xi_c}\kappa\Phi-c_<) \right]
\ee
with $\phi^2<\xi^{-1}_c$ and
\be{3.38}
\phi=\frac{1}{\sqrt{\xi_c}} \left[ (\coth(\sqrt{\xi_c}\kappa\Phi-c_>) \right]
\ee
with $\phi^2>\xi^{-1}_c$ respectively.

The conformal factor is according to Eqs. (\ref{3.15}) and (\ref{3.27})
given by
\be{3.39}
\omega=\frac{1}{D-2}\ln(\kappa^2 \vert 1-\xi_c\phi^2 \vert)+A.
\ee
\section{\bf Trapped Internal Dimensions}
\setcounter{equation}{0}
In the following we want to compare the solutions of the minimal model
to those of the corresponding conformal model.
We specify the geometry for the minimal model to be of MCM
type (2.1), with all $M_i$ Ricci flat
(when necessary, assumed to be compact),
hence $R^{(i)}=0$ for $i=1,\ldots,n$.
The minimally coupled scalar field is assumed to have zero potential
$U\equiv 0$.
In the harmonic time gauge (\ref{2.27}) with harmonic time
\begin{equation}
\tau\equiv t^{(m)}_h,
\end{equation}
we demand this model to be a solution for Eq. (\ref{e2.16})
with vanishing $R^{(i)}$ and $U(\Phi)$.
We set $\hat\beta^{n+1}:= \kappa \Phi$
and obtain as solution
a multidimensional (Kasner like) universe,
given by
\be{3.41}
\hat\beta^i=b^i\tau+c^i  \ \mbox{and}\
\hat\gamma=\sum_{i=1}^n d_i \hat\beta^i
=(\sum_{i=1}^n d_i b^i)\tau+(\sum_{i=1}^n d_i c^i),
\ee
with $i=1,\ldots,n+1$,
where with $V\equiv 0$ the constraint Eq. (\ref{e2.17}) simply
reads
\be{3.42}
G_{ij} b^i b^j + (b^{n+1})^2=0.
\ee

With Eq. (\ref{3.39}) the scaling powers of the universe
given by Eqs. (\ref{3.41}) with $i=1,\ldots,n$ transform to corresponding
scale factors of the conformal universe
$$
\beta^i=\hat\beta^i-\omega
$$
\begin{equation}
=b^i\tau+\frac{1}{2-D}\ln\vert 1-\xi_c(\phi)^2\vert
+ c^i + \frac{2}{2-D} \ln\kappa-A
\end{equation}
and
$$
\gamma=\sum_{i=1}^n d_i \hat\beta^i-\omega
$$
\begin{equation}
=(\sum_i d_i b^i)\tau+\frac{1}{2-D}\ln\vert 1-\xi_c(\phi)^2\vert
+(\sum_i d_i c^i) + \frac{2}{2-D} \ln\kappa-A.
\end{equation}
It should be clear that
the variable $\tau$, when harmonic in the minimal model,
in the conformal model cannot be expected to be harmonic either,
i.e. in general
\begin{equation}
\tau\neq t^{(c)}_h.
\end{equation}
Actually from
$$
\gamma=\sum_{i=1}^n d_i \beta^i=\sum_{i=1}^n d_i \hat\beta^i-\omega(D-1)
$$
we see that $\tau= t^{(c)}_h$ only for $D=2$ (but we have $D>2$ !).

Let us take for simplicity
\begin{equation}
A=\frac{2}{2-D} \ln\kappa,
\end{equation}
which yields the lapse function
\be{3.47}
e^\gamma=e^{(\sum_i d_i b^i)\tau+(\sum_i d_i c^i)}
\vert 1-\xi_c(\phi)^2\vert^{\frac{1}{2-D}}
\ee
and for $i=1,\ldots,n$ the scale factors
\be{3.48}
e^{\beta^i}= e^{b^i\tau+ c^i}
\vert 1-\xi_c(\phi)^2\vert^{\frac{1}{2-D}}.
\ee
Let us further set for simplicity
\begin{equation}
c_{<}=c_{>}=\sqrt{\xi_c} c^{n+1}.
\end{equation}

By Eqs. (\ref{3.37})
or (\ref{3.38}),
the minimally coupled scalar field
\begin{equation}
\kappa\Phi(\tau)=b^{n+1}\tau+c^{n+1},
\end{equation}
substituted into Eqs. (\ref{3.47}) and (\ref{3.48}), yields
\be{3.51}
e^\gamma=e^{(\sum_i d_i b^i)\tau+(\sum_i d_i c^i)}
\cosh^{\frac{2}{D-2}}( \sqrt{\xi_c} b^{n+1}\tau )
\ee
resp.
\be{3.52}
e^\gamma=e^{(\sum_i d_i b^i)\tau+(\sum_i d_i c^i)}
\vert\sinh^{\frac{2}{D-2}}( \sqrt{\xi_c} b^{n+1}\tau )\vert
\ee
and, with $i=1,\ldots,n$, non-singular scale factors
\be{3.53}
e^{\beta^i}= e^{b^i\tau+ c^i}
\cosh^{\frac{2}{D-2}}( \sqrt{\xi_c} b^{n+1}\tau )
\ee
resp. singular scale factors
\be{3.54}
e^{\beta^i}= e^{b^i\tau+ c^i}
\vert\sinh^{\frac{2}{D-2}}( \sqrt{\xi_c} b^{n+1}\tau )\vert
\ee
for the conformal model. The scale factor singularity
of the minimal coupling model for $\tau\to-\infty$ vanishes in the conformal
model of Eqs. (\ref{3.51}) and (\ref{3.53}) for a scalar field $\phi$
bounded according to (\ref{3.37}).
For $D=4$ this result had already been indicated by \cite{Ga}.

On the other hand in the conformal model
of Eqs. (\ref{3.52}) and (\ref{3.54}),
with $\phi$ according to (\ref{3.38}),
though the scale factor singularity of the minimal model
for $\tau \to -\infty$ has also disappeared, instead there is another new
scale factor singularity at finite
(harmonic) time $\tau=0$.

Let us consider a special case of the non-singular solution
with $\phi^2<\xi_c^{-1}$,
where we assume the internal spaces
to be static in the minimal model, i.e.  $b^i=0$ for $i=2,\ldots,n$.
Then in the conformal model, the internal spaces are no longer
static. Their scale factors (\ref{3.53}) with $i>2$ have a minimum at
$\tau=0$.
{}From Eq. (\ref{3.42}) with $G_{11}=d_1(1-d_1)$ we find
\begin{equation}
(b^{n+1})^2= d_1 (d_1-1) (b^1)^2.
\end{equation}

With real $b_1$ then also
\be{3.56}
b^{n+1}=\pm\sqrt{d_1 (d_1-1)}b^1
\ee
is real and by Eq. (\ref{3.53}) the scale $a_1$ of $M_1$ has a minimum at
\be{3.57}
\tau_0
=(\sqrt{\xi_c} b^{n+1})^{-1}\artanh\left(
\frac{(2-D)}{2\sqrt{\xi_c}}
\frac{b^1}{b^{n+1}} \right),
\ee
with $\tau_0>0$ for $b^1<0$ and $\tau_0<0$ for $b^1>0$.

Let $M_1$ be the external space with $b^1>0$ and hence $\tau_0<0$.
Let us start with an Euclidean region of complex geometry
given by scale factors
$$
a_k=e^{-ib^k\tau+\tilde c^k}
\vert\sin ( \sqrt{\xi_c} b^{n+1}\tau )\vert^{\frac{2}{D-2}}.
$$
Then we can perform an analytic continuation to the Lorentzian region
with $\tau\to i\tau+\pi/(2 \sqrt{\xi_c} b^{n+1})$, and we require
$c^k=\tilde c^k-i\pi b^k/(2 \sqrt{\xi_c} b^{n+1})$ to be the real
constants of the real geometry.

The quantum creation (via tunnelling)
of different factor spaces takes place at different values of $\tau$
(see Fig. 1).
%
First the factor space $M_1$ comes into real existence and after an
time interval $\Delta\tau=\vert\tau_0\vert$ the internal factor
spaces $M_2, \ldots, M_n$ appear in the Lorentzian region. Since
$\Delta\tau$ may be arbitrarily large, there is in principle an alternative
explanation of the unobservable extra dimensions, independent
from concepts of compactification and shrinking to a fundamental length.
Similar to the spirit of the
idea that internal dimensions might be hidden
due to a potential barrier (\cite{RS}-\cite{Am}),
they may have been up to now still
in the Euclidean region and hence unobservable. This view is also
compatible with their interpretation as complex resolutions of
simple singularities in external space \cite{bRa}.

Now let us perform a transition from Lorentzian time $\tau$ to Euclidean
time $-i\tau$. Then with a simultaneous transition from $b^k$ to
$i b^k$
for $k=1,\ldots,n$ the geometry remains real, since
$\hat\beta^k=b^k\tau+c^k$ is unchanged.
The analogue of Eq. (\ref{3.56}) for the Euclidean region
then becomes
\be{3.58}
b^{n+1}=\pm i\sqrt{d_1 (d_1-1)}b^1.
\ee
This solution corresponds to a classical (instanton) wormhole.
The sizes of the wormhole throats in the factor spaces $M_2,\ldots,M_n$
coincide with the sizes of static spaces in the
minimal model, i.e. $\hat a_2(0),\ldots,\hat a_n(0)$ respectively.

With Eq. (\ref{3.56}) replaced by (\ref{3.58}), the Eq. (\ref{3.57})
remains unchanged in the transition to the Euclidean region,
and the minimum of the scale $a_1$ (unchanged geometry !)
now corresponds to the throat of the wormhole
in the factor space $M_1$.

If one wants to compare the synchronous time pictures of the
minimal and the conformal solution, one has to calculate them for
both metrics.
In the minimal model
we have
\begin{equation}
dt^{(m)}_s=e^{\hat\gamma}d\tau
=e^{(\sum_i d_i b^i)\tau+(\sum_i d_i c^i)}d\tau,
\end{equation}
which can be integrated to
\begin{equation}
t^{(m)}_s=(\sum_i d_i b^i)^{-1}
e^{\hat\gamma}+t_0.
\end{equation}

The latter can be inverted to
\begin{equation}
\tau=(\sum_i d_i b^i)^{-1}
\left\{[\ln(\sum_i d_i b^i)(t^{(m)}_s-t_0)]-(\sum_i d_i c^i)\right\}.
\end{equation}
Setting
\be{3.62}
B:=\sum_{i=1}^{n} d_i b^i \ \mbox{and} \ C:=\sum_{i=1}^{n} d_i c^i,
\ee
this yields the scale factors
\be{3.63}
\hat{a_s}^i=(t^{(m)}_s-t_0)^{b^i/B} e^{\frac{b_i}{B}(\ln B-C)+c_i}
\ee
and the scalar field
\begin{equation}
\kappa\Phi=\frac{b^{n+1}}{B} \{ [\ln B(t^{(m)}_s-t_0)] -C \}+c^{n+1}.
\end{equation}
Let us define for $i=1,\ldots,n+1$ the numbers
\begin{equation}
\alpha^i:=\frac{b^i}{B}.
\end{equation}
With (\ref{3.62}) they satisfy
\be{3.66}
\sum_{i=1}^{n} d_i \alpha^i=1,
\ee
and by Eq. (\ref{3.42}) also
\be{3.67}
\alpha^{n+1}=\sqrt{1-\sum_{i=1}^{n} d_i (\alpha^i)^2}.
\ee
Eqs. (\ref{3.63}) shows, that the solution
(\ref{3.41}) is really a generalized Kasner universe
with exponents $\alpha^i$
satisfying generalized Kasner conditions (\ref{3.66}) and (\ref{3.67}).

In the conformal model the synchronous time is given as
\begin{equation}
t^{(c)}_s=\int e^\gamma d\tau=
\int\cosh^{\frac{2}{D-2}}(\sqrt{\xi_c} b^{n+1}\tau)
e^{B\tau+C} d\tau
\end{equation}
resp.
\begin{equation}
t^{(c)}_s=\int e^\gamma d\tau=
\int \sinh^{\frac{2}{D-2}}(\sqrt{\xi_c} b^{n+1}\tau)
e^{B\tau+C} d\tau.
\end{equation}

\np
\noindent

\section{\bf Conclusion}
\setcounter{equation}{0}
In the first part of this paper,
we reexamine the conformal equivalence between
a model with minimal coupling and one with non-minimal coupling
in the MCM case.
The domains of equivalence are separated by certain
critical values of the scalar field $\phi$.
Furthermore the coupling
constant $\xi$ of the coupling between $\phi$ and $R$ is critical at
both, the minimal value $\xi=0$ and the conformal value
$\xi_c=\frac{D-2}{4(D-1)}$.
In different noncritical regions of $\xi$ a solution of the model
behaves qualitatively very different.

In the second part, we applied the conformal equivalence transformation
to the multidimensional generalized Kasner universe
with minimally coupled scalar field. So we obtained
a new exact solution of a universe with Ricci flat factor spaces
and conformally coupled scalar field.
It has two qualitatively different regions of equivalence:
In the first it is singular w.r.t. scale factors, in the other
it is regular.
For both, static internal spaces in the minimally coupling model
become dynamical in the conformal coupling model.
For the regular solution, scale factors are highly asymmetric
in time.
They have minima at different values harmonic time $\tau$.
These may naturally considered as the different times of birth
of the factor spaces, where they emerge from the classically forbidden
region.
Hence the extra dimensions of the internal factor spaces may be
still trapped, while external space is already born by quantum tunnelling.
In particular it is also possible that some internal spaces never
leave the classically forbidden region.
Analytic continuation of this solution to the Euclidean region
(while pertaining geometry and scalar field real), yields a
classical wormhole (instanton).

\nl\nl
{\Large {\bf Acknowledgements}}
\nl\nl
This work was supported by WIP grant 016659 (U.B.),
in part by DAAD and by DFG grant 436
UKR - 17/7/93 (A. Z.) and DFG grant Bl 365/1-1 (M.R.).
A. Z. also thanks Prof. Kleinert and the Freie Universit\"at Berlin
as well as the members of the Gravitationsprojekt at
Universit\"at Potsdam for their hospitality.

\newpage

\vspace*{10truecm}
{\small Fig. 1: Quantum birth with compact Ricci flat spaces and
birth time $\tau_0\leq 0$ of external Lorentzian space $M_1$. The birth
of internal factor spaces $M_2,\ldots,M_n$ is delayed by the
interval $\Delta\tau=\vert\tau_0\vert$. For $\Delta\tau\to\infty$
the internal spaces remain for ever in the (unobservable) classically
forbidden region.}


\begin{thebibliography}{99}
\bibitem{IMZ}
V. D. Ivashchuk, V. N. Melnikov, A. I. Zhuk, Nuovo Cim. B
{\bf 104}, 575 (1989),
\bibitem{aBlLP}
U. Bleyer, D.-E. Liebscher and A. G. Polnarev,
Nuovo Cim. B {\bf 106}, 107 (1991).
\bibitem{bBlLP}
U. Bleyer, D.-E. Liebscher and A. G. Polnarev,
{\em Kaluza-Klein Models},
Proc. V$^{th}$ Seminar on Quantum Gravity, Moscow 1990, World Scientific
(1991).
\bibitem{IZ}
V. D. Ivashchuk Phys. Lett. A {\bf 170}, 16 (1992).
A. I. Zhuk, Class. Quant. Grav. {\bf 9}, 2029 (1992);
Sov J. Nucl. Phys. {\bf 55}, 149 (1992);
Phys. Rev. D {\bf 45}, 1192 (1992).
\bibitem{aBZ}
U. Bleyer and A. Zhuk,
{\em Multidimensional Integrable
Cosmological Models with Positive External Space Curvature};
{\em Multidimensional Integrable
Cosmological Models with Negative External Space Curvature}.
Gravitation and Cosmology {\bf 1} (1995), in press.
\bibitem{bBZ}
U. Bleyer and A. Zhuk,
Class. Quant. Grav. {\bf 11}, 1 (1994).
\bibitem{cBZ}
U. Bleyer and A. Zhuk,
Nucl. Phys. B {\bf 429}, 177 (1994).
\bibitem{dBZ}
U. Bleyer and A. Zhuk,
{\em Kasner-like and Inflation-like Solutions in
Multidimensional Cosmology, in preparation}.
\bibitem{Ma}
K. Maeda, Phys. Rev. D {\bf 39}, 3159 (1989).
\bibitem{Xa}
B. C. Xanthapoulos and Th. E. Dialynas, J. Math. Phys. {\bf 33}.
1463 (1992).
\bibitem{Pa}
D. N. Page, J. Math. Phys. {\bf 32}, 3427 (1991).
\bibitem{aRa}
M. Rainer, {\em Conformal Coupling and Invariance in Arbitrary
Dimensions}, Int. J. Mod. Phys. D (1994), in press.
\bibitem{RS}
V. A. Rubakov, M. E. Shaposhnikov, Phys. Lett. B {\bf 125}, 136 (1983).
\bibitem{Vis}
M. Visser, Phys. Lett. B {\bf 159}, 22 (1985).
\bibitem{Sq}
E. J. Squires, Phys. Lett. B {\bf 167}, 286 (1986).
\bibitem{Mai}
M. D. Maia and V. Silveira, Phys. Rev. D {\bf 48}, 954 (1993).
\bibitem{Am}
L. Amendola, E. W. Kolb, M. L. Litterio and F. Occhionero,
Phys. Rev. D {\bf 42}, 1944 (1990).
\bibitem{Tr}
E. P. Tryon,
Nature {\bf 246}, 396 (1973).
\bibitem{Fo}
P. I. Fomin,
Doklady A.N.UkrSSR {\bf 9}, 831 (1975).
\bibitem{Zel}
Ya. B. Zeldovich,
Pis'ma Astron. Zh. (Sov. Astron. Lett.) {\bf 7}, 579 (1981).
\bibitem{Gr}
L. P. Grishchuk, Ya. B. Zeldovich,
Proc. II$^{nd}$ Sem. on Quantum Gravity, Moscow (1982).
\bibitem{Vi}
A. Vilenkin,
Phys. Lett. B {\bf 117}, 25 (1982);
Phys. Rev. D {\bf 27}, 2848 (1983).
\bibitem{bRa}
M. Rainer, {\em Projective Geometry for Relativistic Quantum
Physics}, Proc. 23$^{rd}$ Ann. Iranian Math. Conf. (Baktaran, 1992);
J. Math. Phys. {\bf 35}, 646 (1994).
\bibitem{Gi}
G. W. Gibbons and S. W. Hawking,
Phys. Rev. D {\bf 15}, 2752 (1977).
\bibitem{Ga}
D. V. Gal'tsov and B. C. Xanthopoulos, J. Math. Phys. {\bf 33},
273 (1992).

\end{thebibliography}
\end{document}